\documentclass[prb,preprint,showpacs,preprintnumbers,amsmath,amssymb]{revtex4}

\usepackage{graphicx,subfigure}
\usepackage{dcolumn}
\usepackage{bm}
\usepackage{psfrag}
\usepackage{multirow}

\def\be{\begin{equation}}       \def\ee{\end{equation}}
\def\bea{\begin{eqnarray}}      \def\eea{\end{eqnarray}}
\def\half{\frac{1}{2}}
\def\third{\frac 1{3}}
\def\six{\frac 1{6}}
\def\dag{\dagger}
\def\non{\nonumber}
\begin{document}

\title{Zero temperature phases of the frustrated   
J$_1$-J$_2$ antiferromagnetic spin-1/2 Heisenberg model on a simple cubic lattice}
\author{Kingshuk Majumdar}
\affiliation{Department of Physics, Grand Valley State University, Allendale, 
Michigan 49401, USA}
\email{majumdak@gvsu.edu}
\author{Trinanjan Datta}
\affiliation{Department of Chemistry and Physics, Augusta State University, Augusta, 
Georgia 30904, USA}
\email{tdatta@aug.edu}
\date{\today}

\begin{abstract}\label{abstract}
At zero temperature magnetic phases of the 
quantum spin-$1/2$ Heisenberg antiferromagnet on a simple 
cubic lattice with competing first and second neighbor exchanges ($J_1$ and $J_2$) is 
investigated using the non-linear spin wave theory. We find 
existence of two phases: a two sublattice N\'{e}el phase for small $J_2$ (AF),
and a collinear antiferromagnetic phase at large $J_2$ (CAF).  We obtain the sublattice 
magnetizations and ground state energies for the two phases and find that there exists
a first order phase transition from the AF-phase to the CAF-phase at the critical 
transition point, $p_c=0.28$. Our results for the value of $p_c$ are in excellent agreement with 
results from Monte-Carlo simulations and variational spin wave theory. We also show that
the quartic $1/S$ corrections due spin-wave interactions enhance the sublattice magnetization 
in both the phases which causes the intermediate paramagnetic phase predicted from linear
spin wave theory to disappear.
\end{abstract}

\maketitle
\section{Introduction}
Frustrated quantum Heisenberg magnets with competing nearest 
neighbor (NN) and next-nearest-neighbor (NNN) antiferromagnet (AF) exchange interactions, 
$J_1$ and $J_2$ respectively, have been under intense investigation both theoretically and 
experimentally in 
condensed matter physics for more than a decade.~\cite{diep1} At low temperatures these 
systems exhibit new types of magnetic order and novel quantum phases.~\cite{diep1,subir1,subir2} 
A well-known example is the quantum spin-$1/2$ antiferromagnetic $J_1-J_2$
model on a square lattice, which has been studied extensively by various analytical
and numerical methods.~\cite{tassi1,tassi2,oleg,subir3,dot,chubu1,gelfand,sushkov,weihong,
singh,irkhin} For this two-dimensional square lattice system
with $J_2=0$ the ground state is antiferromagnetically
ordered at zero temperature. Addition of next nearest neighbor interactions induces
a strong frustration and break the antiferromagnetic (AF) order. The competition between the 
NN and NNN interactions for the square lattice is characterized by the frustration parameter 
$p= J_2/J_1$. It has been found that a disordered quantum spin liquid phase exists between 
$p_{1c} \approx 0.38$ and $p_{2c} \approx 0.60$.
For $p<p_{1c}$ the square lattice is AF-ordered whereas for $p>p_{2c}$ a collinear phase
emerges. In the collinear state the NN spins have a parallel orientation in the vertical
direction and antiparallel orientation in the horizontal direction or vice versa. The nature of
phase transition from AF-ordered state to disordered state at $p_{1c}$ is of second order 
and from the disordered state to the collinear state at $p_{2c}$ is of first order.

The properties of quantum magnets depend strongly on the lattice dimensionality since the 
tendency to order is more pronounced in three dimensional (3D) systems than in the lower 
dimensional systems. Furthermore, in 3D the available phase space is more and we expect 
quantum fluctuations to play a lesser role as compared to 1D and 2D. In 1D and 2D the 
available phase space is limited and quantum fluctuations play a dominant role in determining 
the quantum critical points. Despite this fact a magnetically disordered phase has been observed 
in frustrated 3D systems such as the Heisenberg AF on the pyrochlore 
lattice~\cite{canals} or on the stacked kagome lattice~\cite{Fak08SL,subir92,chubu92Kagome,
harris92Kagome}. Studies 
on the Heisenberg AF on the pyrochlore 
lattice (a geometrically frustrated system) have revealed the existence of a spin liquid 
state.~\cite{canals} On the other hand, for the 3D J1-J2 model on the body-centered cubic (BCC) lattice 
there are no signs of an intermediate quantum paramagnetic phase at zero 
temperature.~\cite{oitmaa,Schmidt,kingBCC} For the BCC 
lattice competing interactions and not lattice geometry generates the frustration. This comparison 
illustrates how the magnetic phase diagram may dramatically change based on whether the 
frustration is generated by \emph{competing interactions} or \emph{by geometry}.

Most of the efforts on quantum 3D magnets have primarily focused on geometrically 
frustrated lattices.~\cite{diep1,subir1} There exists some computational\cite{azaria,derrida1,
derrida2,lallemand,karchev1,karchev2,banavar,oitmaa,Schmidt,oguchi,katanin,diep4,ader} 
and very few 
analytical studies\cite{villain1,villain2,banavar,kingBCC} of the magnetic phase diagrams and magnetic order of spin-$1/2$ 
Heisenberg AF on 3D lattices where on the study of 
magnetic phase diagrams and magnetic order of spin-$1/2$ Heisenberg AF on 3D lattices 
where \emph{competing interactions induce frustration}.~\cite{villain1,villain2,azaria,derrida1,
derrida2,lallemand,karchev1,karchev2,banavar,oitmaa,Schmidt,kingBCC,oguchi,katanin,diep4,ader}

Very few analytical and numerical results exist for the the frustrated $J_1 - J_2$ isotropic
Heisenberg model on a simple cubic lattice.~\cite{diep2,diep3,irkhin,kishi} This model has
been studied previously using Monte Carlo simulation~\cite{diep2}, variational spin wave 
theory~\cite{kishi}, and modified spin wave theory~\cite{irkhin}. In a recent work 
the critical 
properties of the 3D anisotropic quantum spin-$1/2$ model on a 
simple cubic (SC) lattice has been investigated within the framework of the differential 
operator technique and by using an effective field theory in a 
two-spin cluster.~\cite{viana} The study revealed that at zero temperature there is a 
AF-lamellar (first order) phase transition. 
The motivation for the present work is to investigate the zero temperature phases 
of this model in the framework of non-linear spin wave theory (NLSWT) and to obtain the critical
transition points of this model. Also we will compare our results from NLSWT with the prediction
from the linear spin wave theory (LSWT). 

The paper is organized as follows. In Section~\ref{sec:model} we set-up the Hamiltonian 
for the spin-$1/2$ Heisenberg AF on the SC lattice. The classical ground state
configurations of the model and the different phases are then discussed. 
In Section~\ref{sec:SCNLSWT} we map the spin Hamiltonian to the Hamiltonian of interacting 
bosons and 
develop the NLSWT sublattice magnetization and energy expressions. The sublattice 
magnetizations and the ground state energies for the two phases are numerically 
calculated and the results are plotted and discussed in Section~\ref{sec:results}. 
Finally we summarize our findings in Section~\ref{sec:conclusion}.  

\section{Classical Ground State Configurations \label{sec:model}}
The Hamiltonian for a spin-$1/2$ Heisenberg antiferromagnet  
with first and second neighbor interactions on a simple cubic (SC) lattice is 
\be
H = \half J_1 \sum_{\langle ij \rangle}{\bf S}_i \cdot {\bf S}_j 
+ \half J_2 \sum_{[ij]}{\bf S}_i \cdot {\bf S}_j.
\label{ham}
\ee
$J_1$ is the NN and $J_2$ is the frustrating NNN
(which are along the face diagonals of the cube) exchange constants. Both couplings are considered 
antiferromagnetic, 
i.e. $J_1,J_2>0$. For the SC lattice the number of nearest and next-nearest neighbors are
$z_1=6$ and $z_2=12$.

The limit of infinite spin, $S \rightarrow \infty$, corresponds to the classical 
Heisenberg model. We assume that classically the 
spin configurations of the system are described by 
$S_i=S{\bf u}e^{i{\bf q}\cdot{\bf r_i}}$,
where ${\bf u}$ is a vector expressed in terms of an arbitrary orthonormal basis and
${\bf q}$ defines the relative orientation of the spins on the lattice.
The classical ground state energy of 
the lattice in terms of the frustration parameter, p, is given by
\be
E_{\bf k}/NJ_1
= \half S^2z_1[\gamma_{1 {\bf k}}+p\gamma_{2 {\bf k}}], 
\ee
with the structure factors
\begin{eqnarray}
\gamma_{1{\bf k}} &=& \frac 1{3}\Big[ \cos(k_x)+\cos(k_y)+\cos(k_z)\Big], \\
\gamma_{2{\bf k}} &=& \frac 1{3}\Big[\cos(k_x)\cos(k_y)+\cos(k_y)\cos(k_z) 
+\cos(k_z)\cos(k_x)\Big],
\end{eqnarray}
where we define the parameter of frustration as $p=z_2J_2/z_1J_1$.~\cite{note1}
The wave-vectors along the $x$, $y$, and $z$ directions are denoted by $k_x,k_y$, and $k_z$. 
The number of lattice sites are given by $N$ and we have set the lattice spacing $a=1$.

At zero temperature the classical ground state (GS) for the SC lattice can be 
characterized by the values of $p$. $p=0$ corresponds to the unfrustrated case (only AF 
interactions between NN). For $p<1/2$ or $J_2/J_1 <1/4$, there is a single minimum in energy
$E_0/NJ_1=-3S^2(1-p)$ for the wave-vector 
$(\pm \pi,\pm \pi,\pm \pi)$. They correspond to the classical two sublattice 
N\'{e}el state (AF phase) where spins in A and B-sublattices point in opposite 
directions [Fig.~\ref{fig:phase1}].

For $p>1/2$ apart from 
the global rotation the classical ground state has an infinite degeneracy -- the frustration
is uniformly distributed on all the spins, causing a non-collinear GS with very large
degeneracy. In general, the GS for $p>1/2$ can be decomposed into two NNN tetrahedra. The
spin configurations in each of these two NNN tetrahedra can be characterized by two angles
$\theta$ and $\phi$. This results in a four sublattice $A,B,C,D$ [Fig.~\ref{fig:phase2}] 
antiferromagnetic structure. Out of these infinite possibilities there are three collinear 
configurations (one line up, one line down). The wave-vectors
corresponding to these collinear states are $(\pi,\pi,0),\;(\pi,0,\pi),\;(0,\pi,\pi)$.
The classical GS energy for these states is $E_0/NJ_1=-S^2(1+p)$.
Thermal or quantum fluctuations lift these degeneracies and select specific 
discrete states and it has been conjectured that thermal or quantum disorder favors 
collinear states ({\em order by disorder}).~\cite{shender,henley} The four fold 
rotational symmetry of the lattice is spontaneously broken in this state. By 
employing a spin wave theory based on the general four sublattice mean field
ground state it has been shown that the quantum fluctuations stabilize a collinear
spin ordering.~\cite{kubo} Quantum Monte Carlo 
simulations on the frustrated SC lattice for $p>1/2$ also confirm this conjecture.~\cite{diep2} 
In the present article, for $p>1/2$, we consider the system to be in one of these three 
collinear configurations (collinear antiferromagnet or CAF).

$p=1/2$ corresponds to the case where both $J_1$ and $J_2$ compete -- causing frustration in the 
system. This critical value $p_{\rm class}=0.5$ is the classical phase transition point where 
a phase transition from AF to CAF phase occurs. In this work we will investigate the role of
quantum fluctuations in the two different phases (AF and CAF) of the model and how these fluctuations shift
the critical transition point.

\section{\label{sec:SCNLSWT} Self-consistent non-linear spin wave theory}
The Hamiltonian in Eq.~\ref{ham} can be mapped into an equivalent Hamiltonian
of interacting bosons by transforming the spin operators to bosonic operators 
$a, a^\dag$ and $b, b^\dag$ using the 
well-known Holstein-Primakoff transformations. For the AF-phase ($J_2<J_1$) the 
operators $a, a^\dag$ and $b, b^\dag$ are for the $A$ and $B$ sublattices. On the other hand for 
the CAF phase ($J_2>J_1$) we have used the operators $a, a^\dag$ and $b, b^\dag$ for the up 
and down spin configurations. 
\begin{eqnarray}
S_{Ai}^+ &\approx& \sqrt{2S}\Big(1- \frac {a_i^\dag a_i}{4S} \Big)a_i,\;\;
S_{Ai}^- \approx\sqrt{2S}a_i^\dag \Big(1-\frac {a_i^\dag a_i}{4S}  \Big), \non \\
S_{Ai}^z &=& S-a^\dag_ia_i,  \non \\ 
S_{Bj}^+ &\approx&\sqrt{2S}b_j^\dag \Big(1-\frac {b_j^\dag b_j}{4S}\Big),\;\;
S_{Bj}^- \approx\sqrt{2S}\Big(1-\frac {b_j^\dag b_j}{4S}  \Big)b_j, \non \\
S_{Bj}^z &=& -S+b^\dag_jb_j,\label{spinmap}
\end{eqnarray}
In these transformations we only kept terms up to the order of $1/S$. Next using the
Fourier transforms
\[
a_i = \sqrt{\frac 2{N}}\sum_{\bf k} e^{-i{\bf k \cdot R_i}}a_{\bf k},\;\;\;
b_j = \sqrt{\frac 2{N}}\sum_{\bf k} e^{-i{\bf k \cdot R_j}}b_{\bf k},
\]
the real space Hamiltonian is transformed to the ${\bf k}$-space Hamiltonian.
In the following two
sections we study the cases $J_2<J_1$ and $J_2>J_1$ separately.

\subsubsection{\label{sec: smallp}$J_2<J_1$: AF phase}
In this phase the classical ground state is the two-sublattice N\'{e}el 
state [Fig.~\ref{fig:phase1}]. For
the NN interaction, spins in $A$ sublattice interacts with spins in
$B$ sublattice and vice versa. On the other hand the NNN 
exchange $J_2$ connects spins on the same sublattice, $A$ with $A$ and $B$ with
$B$. Substituting equations (\ref{spinmap}) into (\ref{ham}), the ${\bf k}$-space 
Hamiltonian takes the form:
\be
H=H^{(0)}+H^{(2)}+H^{(4)}.
\ee
The classical ground state energy $H^{(0)}$ and the 
quadratic terms $H^{(2)}$ are
\bea
H^{(0)} &=& -\half NJ_1S^2 z_1 (1-p),
\label{cgs}\\
H^{(2)} &=& J_1S z_1 \sum_{\bf k}\Big[A^{(1)}_{0{\bf k}}(a_{\bf k}^\dag a_{\bf k}
 +b_{\bf k}^\dag b_{\bf k}) +
B^{(1)}_{0{\bf k}}(a^\dag_{\bf k}b^\dag_{-\bf k}+a_{-\bf k}b_{\bf k})\Big]. 
\label{smallp}
\eea
with the coefficients $A^{(1)}_{0{\bf k}}$ and $B^{(1)}_{0{\bf k}}$  defined as
\begin{eqnarray}
A^{(1)}_{0{\bf k}}&=&1-p(1-\gamma_{2{\bf k}}), \label{Ak0smallp}\\ 
B^{(1)}_{0{\bf k}}&=& \gamma_{1{\bf k}}.
\label{Bk0smallp}
\end{eqnarray}

The quartic terms in the Hamiltonian $H^{(4)}$ involve interactions between $A-B$
(for NN terms) and $A-A,\; B-B$ (for NNN terms) sublattices. The Hamiltonian for these 
interaction are stated in Appendix~\ref{Smallp}, Eq.~\ref{quartic}.
These terms are evaluated by applying the Hartree-Fock
decoupling process. The contributions of the decoupled quartic terms 
to the harmonic Hamiltonian in Eq.~\ref{smallp} are to redefine the values
of $A^{(1)}_{0{\bf k}}$ and $B^{(1)}_{0{\bf k}}$ which are now
\bea
A^{(1)}_{\bf k} &=& \Big( 1-\frac {u_1+v_1}{S}\Big) 
- p[1-\gamma_{2 {\bf k}}]\Big( 1- \frac {u_1-w_1}{S} \Big),  
\label{Acoeff}\\
B^{(1)}_{\bf k} &=& \gamma_{1{\bf k}}\Big( 1-\frac {u_1+v_1}{S} \Big),
\label{Bcoeff}\\
\omega^{(1)}_{\bf k}&=& \sqrt{\Big(A^{(1)}_{\bf k}\Big)^2-\Big(B^{(1)}_{\bf k}\Big)^2}.
\label{omegak}
\eea 
The coefficients $u_1,v_1$ and $w_1$ are in Appendix~\ref{Smallp}.
They are evaluated self-consistently from equations (\ref{Acoeff}), (\ref{Bcoeff}), 
(\ref{u1})--(\ref{w1}).

The quartic corrections to the ground state energy is calculated from the four-boson
averages. In the leading order they are decoupled into the bilinear combinations
(equations (\ref{u1}) -- (\ref{w1}))
using Wick's theorem. The corresponding four boson 
terms are,
\bea
\langle a^\dag_i a_i b^\dag_j b_j \rangle &=& u_1^2+v_1^2,\;\;\; 
\langle a^\dag_i b^\dag_j b_jb_j \rangle = 2u_1v_1,
 \non \\
\langle a^\dag_i a_i a_i b_j \rangle &=& 2u_1v_1,\;\;\;  
\langle a^\dag_i a_i a_j^\dag a_j \rangle = u_1^2+w_1^2,  \\
\langle a_i a_j^\dag a_j^\dag a_j \rangle &=& 2u_1w_1, \;\;\;
\langle a^\dag_i a_i a_i a_j^\dag \rangle = 2u_1w_1. \non 
\eea 
This yields the ground state energy correction from the quartic terms:
\be
\delta E^{(4)}= -\half NJ_1z_1\big[(u_1+v_1)^2-p(u_1-w_1)^2\big].
\label{EquarticsmallJ}
\ee
Adding all the corrections together the ground state energy takes the form
\bea
E/NJ_1 &=&-\half z_1S(S+1)(1-p)+
\half z_1S \Big[\frac 2{N} \sum_{\bf k} \omega^{(1)}_{\bf k}\Big] \non \\
&+& \half z_1\big[(u_1+v_1)(1-u_1-v_1)- p(u_1-w_1)(1-u_1+w_1)\big]
\label{Esmallp}
\eea
and the average sublattice magnetization $\langle S_\alpha \rangle$ 
is given by 
\be
\langle S_\alpha \rangle = S\Big[1-\frac 1{2S}
\Big\{\frac 2{N} \sum_{\bf k}\frac {A^{(1)}_{\bf k}}{\omega^{(1)}_{\bf k}}
-1\Big\}\Big].
\label{magsmallp}
\ee
Using equations (\ref{Acoeff})--(\ref{omegak}), we numerically evaluate $E/NJ_1$ and
$\langle S_\alpha \rangle$.

\subsubsection{\label{sec: largep}$J_2>J_1$: CAF phase}
The classical ground state for $J_2>J_1$ is considered to be in one of the three collinear
states [Fig.~\ref{fig:phase2}]. For NN and NNN exchanges there are $A-B$, $A-C$, $A-D$, 
$B-C$, $B-D$, and $C-D$ interactions between the four sublattices 
[See Fig.~\ref{fig:phase2}]. Considering all their contributions together up to the quadratic
terms the harmonic Hamiltonian takes the same form as before with
\bea
H^{(0)} &=& -\frac {1}{6} NJ_1S^2z_1(1+p), \\
A^{(2)}_{0{\bf k}}&=& \third (1+\cos k_z)+\third p(1+\cos k_x \cos k_y),\label{Ak0largep}\\ 
B^{(2)}_{0{\bf k}}&=& \third (\cos k_x + \cos k_y)(1+p\cos k_z).
\label{Bk0largep}
\eea
The quartic terms in the Hamiltonian for this case are shown
in Appendix~\ref{Bigp}. These terms are decoupled and evaluated in the same way as before. The 
renormalized values of the coefficients $A^{(2)}_{\bf k}$ and $B^{(2)}_{\bf k}$ are 
\bea
A^{(2)}_{\bf k} &=& A^{(2)}_{0{\bf k}}
+ \frac 1{3S}\big[({\overline u}-{\overline w_z})(1-\cos k_z)-2({\overline u}+
{\overline v_{1}}) \non \\
&+& p\big\{({\overline u}-{\overline w_1})(1-\cos k_x \cos k_y)-2({\overline u}+
{\overline v_2})\big\}\big],
\label{largeA}\\
B^{(2)}_{\bf k} &=& B^{(2)}_{0{\bf k}}
- \frac 1{3S}\big\{({\overline u}+{\overline v_1})
+ p({\overline u}+{\overline v_2})\cos k_z\big\}(\cos k_x+\cos k_y), 
\label{largeB}\\
\omega^{(2)}_{\bf k}&=&\sqrt{\Big(A^{(2)}_{\bf k}\Big)^2-\Big(B^{(2)}_{\bf k}\Big)^2}.
\label{omega2k}
\eea 
The coefficients ${\overline u},{\overline v_{1}},{\overline v_2},{\overline w_1},
{\overline w_z}$ are in Appendix~\ref{Bigp}.
As before these coefficients are calculated self-consistently from 
equations (\ref{largeA})--(\ref{omega2k}) and (\ref{u2})--(\ref{w2}).
The quartic correction to the ground state energy (following the same Hartree-Fock decoupling
process as done in the AF-case) is
\be
\delta E^{(4)}= -\frac {1}{6}NJ_1z_1\big[2({\overline u}+{\overline v_1})^2
- ({\overline u}-{\overline w_z})^2+2p({\overline u}+{\overline v_2})^2-
p({\overline u}-{\overline w_2})^2\big].
\label{EquarticbigJ}
\ee
Combining all these corrections, the ground state energy takes the following form:
\bea
E/NJ_1 &=& -\frac {1}{6} z_1S(S+1)(1+p)
+\half z_1S \Big[\frac 2{N}\sum_{\bf k} \omega^{(2)}_{\bf k}\Big] \non \\
&-&\six z_1\big[({\overline u}-{\overline w_z})-2({\overline u}+{\overline v_1})
+ p({\overline u}-{\overline w_1})-2p({\overline u}+{\overline v_2})\big]\non \\
&-&\frac {1}{6}z_1\big[2({\overline u}+{\overline v_1})^2
- ({\overline u}-{\overline w_z})^2+2p({\overline u}+{\overline v_2})^2-
p({\overline u}-{\overline w_2})^2\big]. 
\label{Elargep}
\eea

The sublattice magnetization and the ground state energy are then obtained 
numerically using equations (\ref{magsmallp}) and (\ref{Elargep}). 

\section{Results\label{sec:results}}
In Fig.~\ref{fig:param} we show the self-consistent values of the different parameters  
$u_1,v_1,w_1$ (AF phase) and ${\overline u_{1}},{\overline v_{1}},
{\overline v_{2}},{\overline w_{1}},{\overline w_{z}}$ (CAF phase) of our model. These parameters
which provide the quartic corrections to our model do not appear in the LSWT calculations 
for the sublattice magnetization, $\langle S_\alpha \rangle$ and the ground state energy, $E$. 
We see from Fig.~\ref{fig:param} that most of these coefficients vary significantly 
with $p$ especially as $p$ approaches $0.5$ from both ends. This demonstrates that non-linear 
corrections due to the spin-wave interactions play a 
significant role in determining the different phases of our model.

Figure \ref{fig:MagSC} shows the result for the average sublattice magnetization for the SC 
lattice for both AF and CAF phase without (dashed line) and with (solid line) quartic corrections.
In the AF ordered phase or the two sublattice Ne\'el phase where A and B sublattice spins point
in the opposite directions, sublattice magnetization decreases monotonically with increase in $p$ 
until $p \approx 0.49$. This gradual decrease in $\langle S_\alpha \rangle$ with increase in 
$p$ is expected as increasing strength of NNN interaction $J_2$ disorders the antiferromagnetic
spin alignments. With only quadratic terms in the Hamiltonian (linear spin wave theory) we find
that $\langle S_\alpha \rangle$ approaches zero as $p \rightarrow p_{c1}$ where 
$p_{c1} \approx 0.48$ indicating a order-disorder phase transition to a disordered paramagnetic (PM)
state at this point. In the CAF phase with lines of spins up and down, LSWT calculations show that
that $\langle S_\alpha \rangle$ decreases as $p$ approaches 0.5 from above and at $p=p_{c2}=0.50$ 
there is an another phase transition from the CAF state to the disordered PM state.
This is similar to the two dimensional AF-square 
lattice with Heisenberg spins where we have a line of quantum critical points between 
$0.38<p<0.60$. However, self-consistent calculations with quartic $1/S$ corrections 
drastically alter the zero temperature phase diagram. We find that in the AF-phase
with increase in $p$ the system aligns the spins antiferromagnetically along the horizontal
and vertical directions -- thus decreasing the sublattice magnetization 
from $\approx 0.42$ for $p=0$ to 
$\approx 0.30$ for $p=0.49$. In the CAF phase $\langle S_\alpha \rangle$ steadily decreases 
from $\approx 0.41$ for $p=1$ to $\approx 0.27$ for $p=0.52$. There is no existence of any 
disordered state as predicted by the linear spin-wave theory (quadratic corrections).
The disordered PM region disappears completely and we only obtain two phases: AF and CAF. This 
is one of our main findings in the present work. This significant change due to the 
quartic corrections is due to the enhancement of order by quantum fluctuations. 

At $p=0$ (no frustration) there is no quartic corrections to $\langle S_\alpha \rangle$. This can
be observed from equations (\ref{Acoeff}) -- (\ref{omegak}) as the correction factor 
$(1-(u_1+v_1)/S)$ cancels out in equation \ref{magsmallp}. Our non-linear spin wave theory 
calculations become unstable close
to the classical transition point $p_{\rm class}=0.5$ since the coefficient $A_{\bf k}^{1}$ 
becomes equal to $B_{\bf k}^{1}$.

We have also applied the NLSWT technique to compute the quartic corrections in the 
spin-$1/2$ Heisenberg AF on a body-centered
lattice.~\cite{kingBCC} The LSWT calculation for the BCC lattice does not predict any 
intermediate disordered state and the quartic corrections play a role in stabilizing 
the sublattice magnetization (see Fig.~2 of Ref.~\onlinecite{kingBCC}). However, the effect 
of quartic corrections is more pronounced in the SC lattice where the intermediate disordered 
phase disappears.

In Fig.~\ref{fig:GSE} we show the ground state energy per site, $E/NJ_1$, for the AF and 
the CAF phases with (solid line) and without (dashed line) quartic corrections as a function of the 
frustration parameter $p$. Classically $p_{\rm class}=0.5$ or $J_2/J_1=0.25$ 
is the critical point where a phase transition from the AF phase to one of the three CAF phases
occur. With increase in frustration (as $p$ approaches 0.5 from both sides) we expect the GS 
energy to increase as the system goes from an energetically favored ordered state to a more 
disordered state. However, linear spin-wave theory calculation fails to capture this. Especially
when $p$ is close to 0.5 we find a slight downward turn in energy. This has been reported in 
Ref.~\onlinecite{kishi}. On the other hand, NLSWT calculation correctly 
produces the expected energy increase. At $p=0$ the calculated energy with the quartic 
corrections is slightly lower than the energy obtained without the quartic corrections. This small
decrease from the LSWT calculation is due to the ground state energy 
correction, which is negative (as seen in Eq.~\ref{EquarticsmallJ} -- these terms originate from 
the self-energy 
Hartree diagrams). As our spin-wave theory calculation becomes unstable 
in the regime $0.49<p<0.52$ we used a spline fit for the AF-phase energy data points and 
then extrapolated the line so that it intersects the CAF-phase energy line. The extrapolated curve
is shown by dotted lines (color online) in the figure. After extrapolation, we find that the two 
energies meet at $p_c \approx 0.56$ or $J_2/J_1 \approx 0.28$. The symmetries of the two phases
are different: SO(3)/SO(2) for the AF phase and Z$_3 \times$SO(3)/SO(2) for the 
CAF phase.~\cite{diep1} 
Due to the different symmetries of the two phases the transition is of first order. This is 
confirmed by the kink in the energy at $p_c \approx 0.28$. Our obtained value from our 
self-consistent NLSWT calculations for the quantum critical point is 
$J_2/J_1 \approx 0.28$. This is another major finding of our work.

Using the variational spin-wave theory the authors in Ref.~\onlinecite{kishi} obtained an 
upper bound of 0.27 for the ground state energy. However, their variational calculation 
slightly overestimates the value of GS energy. This is quite noticeable at $p=0$, where their 
energy value is higher than the LSWT prediction. We have found the $p=0$ energy value to be 
slightly less than the LSWT prediction. This is due to the quartic corrections explained earlier.
The authors in Ref.~\onlinecite{irkhin} used a modified spin-wave theory based on Dyson-Maleev 
representation of the spins to study this model. They numerically obtained the value of critical 
transition point to be $p_c=0.30$. The other known existing numerical work is by 
Diep et. al.~\cite{diep2} By extensive standard and histogram Monte-Carlo simulations, they 
obtained the transition point to be around 0.26. Our result is in excellent agreement with 
the results obtained from Monte-Carlo and variational spin-wave theory 
calculations (our $p_c$ differs by less than 3.5\% from the variational spin-wave theory 
prediction). 

\section{Conclusion\label{sec:conclusion}}
In this work we have investigated the zero temperature phases of a spin-$1/2$ Heisenberg 
frustrated AF on a SC lattice by considering the quartic $1/S$ corrections due to the spin-wave 
interactions. We have compared our results obtained from NLSWT calculations with the predictions 
from LSWT. It is known that LSWT predicts the existence of three phases: a two sublattice 
N\'{e}el  phase for smaller values of the NNN exchange $J_2$, an intermediate paramagnetic 
phase, and a collinear phase for larger values of $J_2$. At zero temperature there are two 
quantum phase transitions - one from the AF-state to the disordered paramagnetic state and the 
other from the disordered state to one of the three collinear states. Both these transitions 
occur at the quantum critical point $p_c \approx 0.5$ or $J_2/J_1 \approx 0.25$. We have found 
that quartic corrections significantly alter this phase diagram as intermediate 
paramagnetic phase disappears. We find the existence of two phases at zero temperature: a 
two sublattice AF phase for small $J_2$ and a collinear phase phase for large $J_2$.  
With the inclusion of quartic interactions the intermediate paramagnetic phase completely 
disappears. Due to the different symmetries of the two phases (AF and CAF) the transition 
between the two phases is of first order and we find that the critical point for transition 
to be 0.28. Our obtained result is 
in excellent agreement with existing numerical results from Monte Carlo simulations. 

\begin{acknowledgments}
One of us (KM) thanks O. Starykh and S. D. Mahanti for helpful discussions.
\end{acknowledgments}
\appendix
\section{\label{Smallp} Quartic terms for the AF phase}
The quartic terms for the AF phase from the NN interactions
involve $A-B$ interactions and for the NNN
interactions involve $A-A$ and $B-B$ type interactions. Considering all these
interactions the
quartic Hamiltonian takes the form:
\bea
H^{(4)} &=& -J_1 \sum_{\langle ij \rangle}\Big[a_i^\dag a_i b^\dag_j b_j 
+\frac 1{4}\Big(a_ib^\dag_jb_jb_j+a_i^\dag a_ia_ib_j 
+ h.c.\Big)\Big] \non \\
&+& \half J_2\sum_{[ij]}\Big[a^\dag_ia_ia^\dag_ja_j-\frac 1{4}\Big(a_ia^\dag_ja_j^\dag a_j 
+ a_i^\dag a_ia_ia_j^\dag+h.c.\Big) 
+ a \leftrightarrow b\Big]. 
\label{quartic}
\eea
In the harmonic approximation the following
Hartree-Fock averages are non-zero for the SC-lattice Heisenberg 
antiferromagnet:
\bea
u_1 &=& \langle a_i^\dag a_i \rangle = \langle b_i^\dag b_i \rangle = 
\half \Big[\frac 2{N} \sum_{\bf k}
\frac {A^{(1)}_{{\bf k}}}{\omega^{(1)}_{{\bf k}}}
 -1\Big], \label{u1} \\
v_1 &=& \langle a_i b_j \rangle = \langle a_i^\dag b_j^\dag \rangle
=-\half \Big[\frac 2{N} \sum_{\bf k}
\frac {\gamma_{1{\bf k}}B^{(1)}_{{\bf k}}}{\omega^{(1)}_{{\bf k}}}
\Big],\label{v1}\\
w_1 &=& \langle a_i^\dag a_j \rangle = \langle b_i^\dag b_j \rangle =
\half \Big[\frac 2{N} \sum_{\bf k}
\frac {\gamma_{2{\bf k}}A^{(1)}_{{\bf k}}}{\omega^{(1)}_{{\bf k}}}
\Big], \label{w1}
\eea 
where $\omega^{(1)}_{{\bf k}}=\sqrt{\Big(A^{(1)}_{{\bf k}}\Big)^2 -
\Big(B^{(1)}_{{\bf k}}\Big)^2}.$

\section{\label{Bigp} Quartic terms for the CAF phase}
\begin{figure}[httb]
\centering
\subfigure[\;AF phase]
{\includegraphics[width=2.5in]{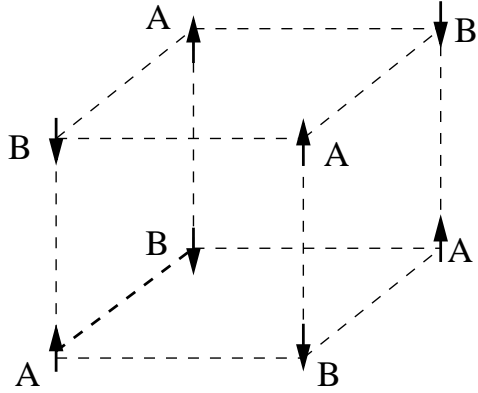}\label{fig:phase1}}
\hfill
\subfigure[\;CAF phase]{
\includegraphics[width=3.0in]{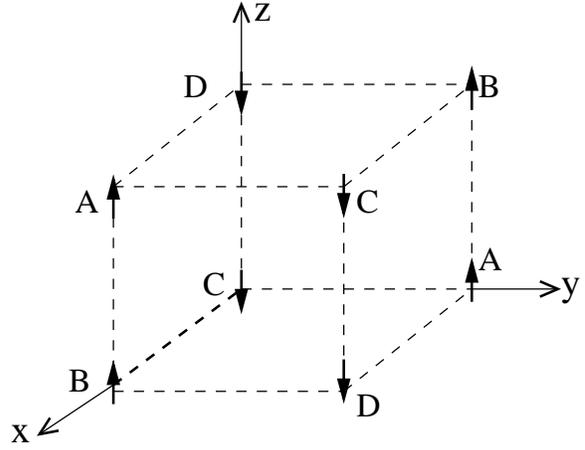}\label{fig:phase2}}
\caption{\label{fig:SCphases} AF and collinear antiferromagnetic (CAF) ordered phases of the 
SC lattice.
In the AF phase all $A$-sublattice spins point in the direction of an 
arbitrary unit vector while $B$-sublattice spins point in the 
opposite direction. For the CAF phase the spin configurations (lines of spins 
up and down) of the four
sublattices A,B,C, and D are shown in the Figure. There are two other equivalent 
configurations 
with lines along the two other directions of the cubic lattice.}
\end{figure}

\begin{figure}[httb]
\centering
\includegraphics[width=5in]{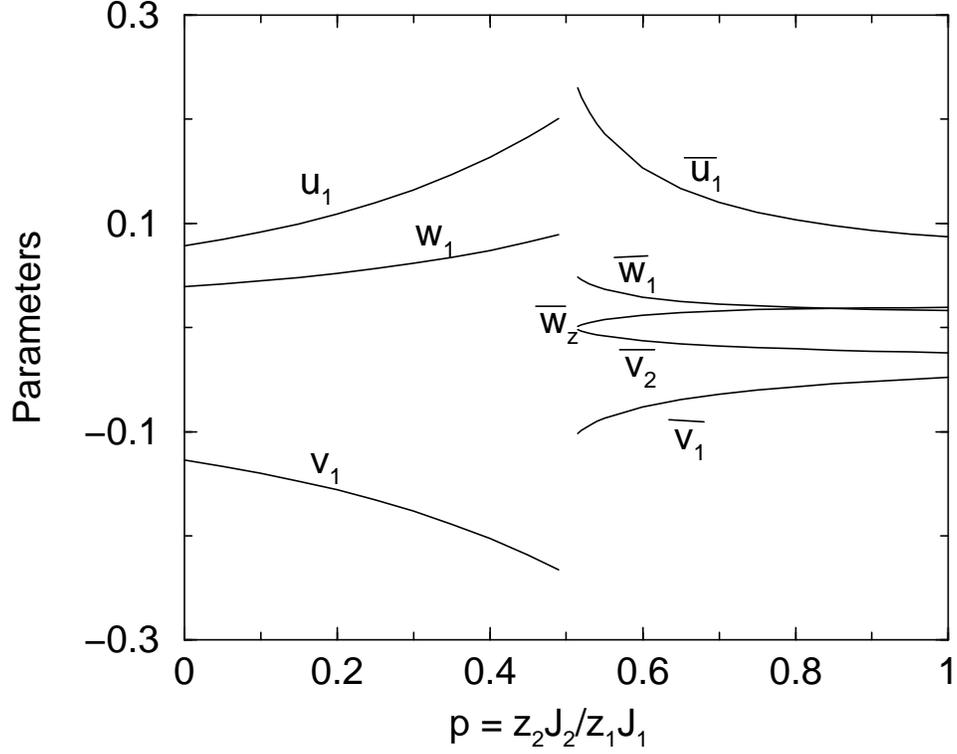}
\caption{\label{fig:param} Self-consistent results for the different parameters, 
$u_1,v_1,w_1$ (for the AF-phase) and ${\overline u_{1}},{\overline v_{1}},
{\overline v_{2}},{\overline w_{1}},{\overline w_{z}}$ (for the CAF phase) are plotted with the
frustration parameter $p=z_2J_2/z_1J_1$ (for the SC lattice $z_1=6$ and $z_2=12$). These 
coefficients vary significantly with $p$, which
shows that the quartic interaction terms play a significant role in determining the
different phases of our model.}
\end{figure}

\begin{figure}[httb]
\centering
\includegraphics[width=5in]{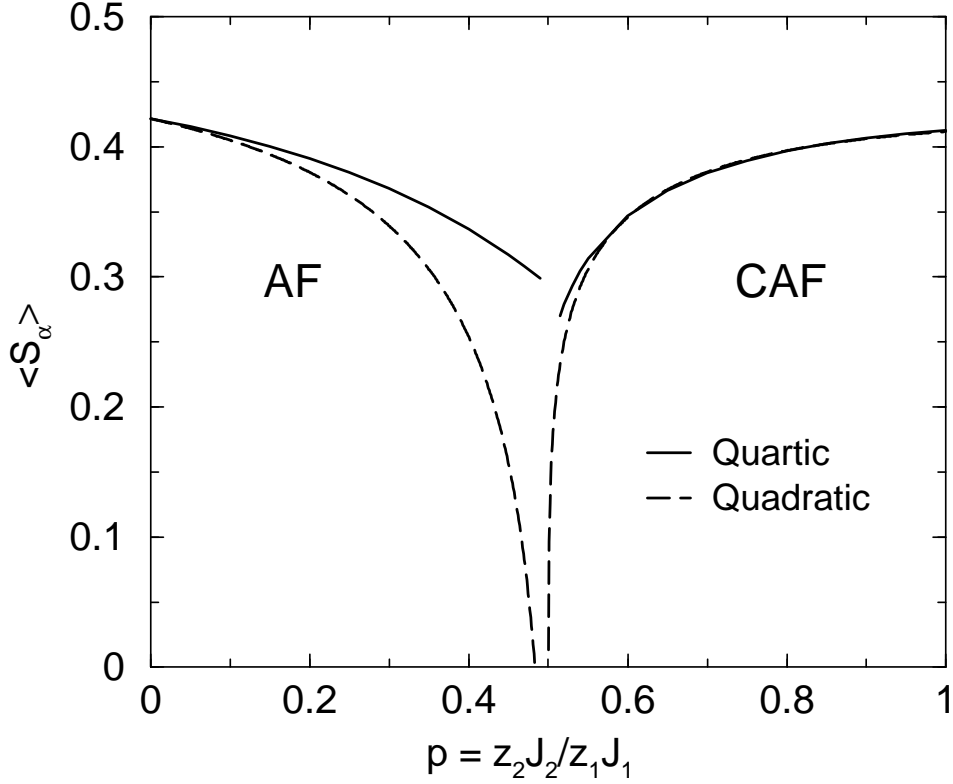}
\caption{\label{fig:MagSC} Average sublattice magnetization, $\langle S_\alpha \rangle $, 
is plotted with the frustration parameter $p$ for AF and one of the three 
CAF phases with (solid lines) and without (dashed lines) quartic corrections. 
At zero 
temperature without the quartic $1/S$ corrections (linear spin-wave theory) 
$\langle S_\alpha \rangle \rightarrow 0$ 
at $p_{c1} \approx 0.48$ indicating a phase
transition from the AF-ordered state to the disordered paramagnetic state. At $p_{c2}=0.50$ 
there is a second phase transition from the collinear state to the disordered state for $T=0$. 
Non-linear spin wave theory provides significant corrections to this phase diagram. With the
quartic $1/S$ corrections the disordered PM region disappears completely
and we only obtain two phases: AF and CAF. There is no existence of any disordered state as 
predicted by the linear spin-wave theory (quadratic corrections). 
For both the phases the quartic corrections to the Hamiltonian enhance the magnetic order.}
\end{figure}

\begin{figure}[httb]
\centering
\includegraphics[width=5in]{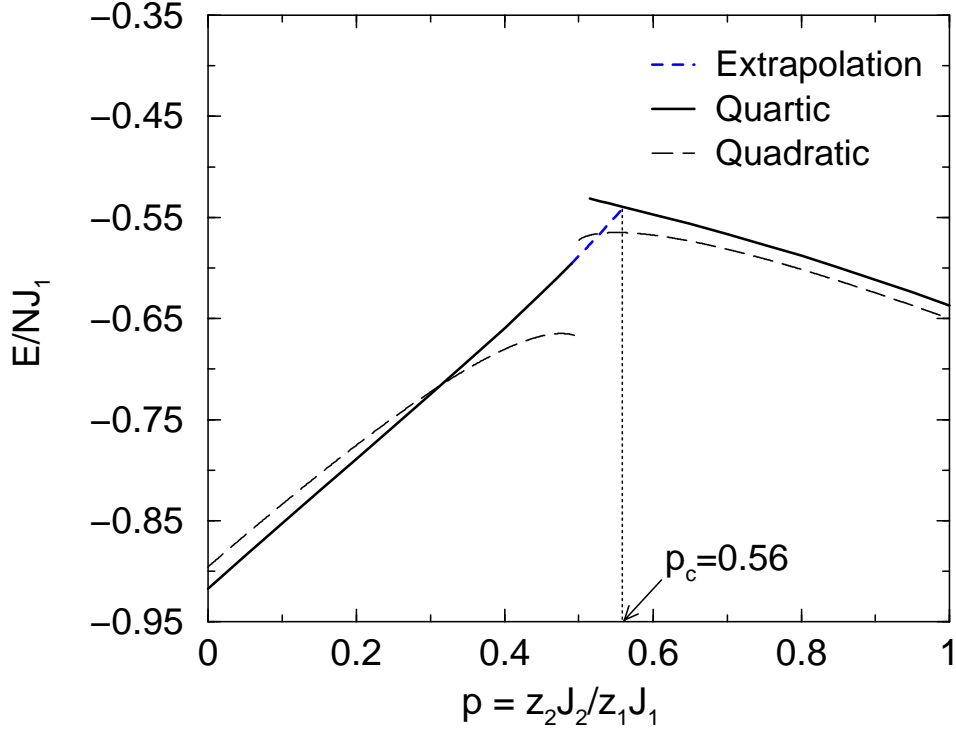}
\caption{\label{fig:GSE} Zero temperature ground state energy per site, $E/NJ_1$,  
is plotted versus $p$  without (dashed lines) and with (solid lines) quartic corrections for both AF ($p<0.5$) and CAF ($p>0.5$) 
ordered phases. Spin wave theory becomes unstable close to the classical 
transition point ($p_{\rm class} \approx 0.5$) between the two phases. After extrapolation 
(shown by the dotted line), we find that the two energies meet at the quantum transition
point, $p_c \approx 0.56$ or 
$J_2/J_1 \approx 0.28$. This kink in the energy indicates a first-order quantum phase transition
from the AF to CAF phase. Compared to the results without quartic corrections (long dashed lines)
we find that the quartic corrections provide significant corrections to the ground state energy
especially near the AF-CAF phase transition point.}
\end{figure}

The quartic terms for the collinear phase from the NN interactions
involve interactions  between the sublattices $B-C$ and $A-D$ along the 
$x$ axis, $A-C$ and $B-D$ along the $y$ axis, and $A-B$ and $C-D$ along the $z$ axis. 
For the NNN interactions spin-spin interactions are between sublattices
$A-B$ and $C-D$ (in the $xy$-plane),  $A-C$ and $B-D$ (in the $xz$-plane), and 
$A-D$ and $B-C$ (in the $yz$-plane) as shown in Fig.~\ref{fig:phase2}.
Adding all the contributions together yield
\bea
H^{(4)} &=&-J_1 \sum_{\langle ij_x \rangle}\Big[a_i^\dag a_i b^\dag_{j_x} b_{j_x} 
+\frac 1{4}\Big(a_ib^\dag_{j_x}b_{j_x}b_{j_x}+a_i^\dag a_ia_ib_{j_x} 
+ h.c.\Big)\Big] \non \\
&-& J_1 \sum_{\langle ij_y \rangle}\Big[a_i^\dag a_i b^\dag_{j_y} b_{j_y} 
+\frac 1{4}\Big(a_ib^\dag_{j_y}b_{j_y}b_{j_y} 
+a_i^\dag a_ia_ib_{j_y} + h.c.\Big)\Big] \non \\
&+& \half J_1\sum_{\langle ij_z \rangle}\Big[a^\dag_ia_ia^\dag_{j_z}a_{j_z}
-\frac 1{4}\Big(a_ia^\dag_{j_z}a_{j_z}^\dag a_{j_z} + a_i^\dag a_ia_ia_{j_z}^\dag 
+ h.c.\Big)+ a \leftrightarrow b\Big]\non \\
&+&\half J_2\sum_{\langle ij_{xy} \rangle}\Big[a^\dag_ia_ia^\dag_{j_{xy}}a_{j_{xy}}
-\frac 1{4}\Big(a_ia^\dag_{j_{xy}}a_{j_{xy}}^\dag a_{j_{xy}} + a_i^\dag a_ia_ia_{j_{xy}}^\dag 
+ h.c.\Big)+ a \leftrightarrow b\Big]\non \\
&-& J_2 \sum_{\langle ij_{yz} \rangle}\Big[a_i^\dag a_i b^\dag_{j_{yz}} b_{j_{yz}} 
+\frac 1{4}\Big(a_ib^\dag_{j_{yz}}b_{j_{yz}}b_{j_{yz}} 
+ a_i^\dag a_ia_ib_{j_{yz}} + h.c.\Big)\Big] \non \\
&-& J_2 \sum_{\langle ij_{xz} \rangle}\Big[a_i^\dag a_i b^\dag_{j_{xz}} b_{j_{xz}} 
+ \frac 1{4}\Big(a_ib^\dag_{j_{xz}}b_{j_{xz}}b_{j_{xz}}+a_i^\dag a_ia_ib_{j_{xz}} 
+ h.c.\Big)\Big].
\label{quartlargeJ}
\eea 
Above $j_x,j_y,j_z$ are NN lattice sites along $x,y,z$ axes 
and $j_{xy},j_{yz},j_{xz}$ connects one lattice site with a NNN corner lattice sites on the 
$xy,yz,xz$ planes.
The different coefficients that originate from Hartree-Fock decoupling process are
\bea
{\overline u} &=& \langle a_i^\dag a_i \rangle = \langle b_i^\dag b_i \rangle = 
\half \Big[\frac 2{N} \sum_{\bf k}
\frac {A^{(2)}_{{\bf k}}}{\omega^{(2)}_{{\bf k}}}
 -1\Big], \label{u2} \\
{\overline  v_{1x}} &=& \langle a_i b_{j_x} \rangle = \langle a_i^\dag b_{j_x}^\dag \rangle
=-\six \Big[\frac 2{N} \sum_{\bf k}
\frac {\cos k_x B^{(2)}_{{\bf k}}}{\omega^{(2)}_{{\bf k}}}
\Big],\\
{\overline v_{1y}} &=& \langle a_i b_{j_y} \rangle = \langle a_i^\dag b_{j_y}^\dag \rangle
=-\six \Big[\frac 2{N} \sum_{\bf k}
\frac {\cos k_y B^{(2)}_{{\bf k}}}{\omega^{(2)}_{{\bf k}}}
 \Big],\\
{\overline v_{2yz}} &=& \langle a_i b_{j_{yz}} \rangle = \langle a_i^\dag b_{j_{yz}}^\dag \rangle
=-\six \Big[\frac 2{N} \sum_{\bf k}
\frac {\cos k_y \cos k_z B^{(2)}_{{\bf k}}}{\omega^{(2)}_{{\bf k}}}
\Big],\\
{\overline v_{2xz}} &=& \langle a_i b_{j_{xz}} \rangle = \langle a_i^\dag b_{j_{xz}}^\dag \rangle
=-\six \Big[\frac 2{N} \sum_{\bf k}
\frac {\cos k_x \cos k_z B^{(2)}_{{\bf k}}}{\omega^{(2)}_{{\bf k}}}
 \Big],\\
{\overline w_z} &=& \langle a_i^\dag a_{j_z} \rangle = \langle b_i^\dag b_{j_z} \rangle =
\six \Big[\frac 2{N} \sum_{\bf k}
\frac {\cos k_z A^{(2)}_{{\bf k}}}{\omega^{(2)}_{{\bf k}}}
\Big],  \\
{\overline w_1} &=& \langle a_i^\dag a_{j_{xy}} \rangle = \langle b_i^\dag b_{j_{xy}} \rangle =
\six \Big[\frac 2{N} \sum_{\bf k}
\frac {\cos k_x \cos k_y A^{(2)}_{{\bf k}}}{\omega^{(2)}_{{\bf k}}}
\Big], \label{w2} 
\eea
where $\omega^{(2)}_{{\bf k}}=\sqrt{\Big(A^{(2)}_{{\bf k}}\Big)^2 -\Big(B^{(2)}_{{\bf k}}\Big)^2}.$
By symmetry ${\overline  v_{1x}}={\overline  v_{1y}}={\overline v_1}$ and
${\overline  v_{2yz}}={\overline  v_{2xz}}={\overline v_2}$.
\bibliography{SC}

\begin{thebibliography}{43}
\bibitem[{not()}]{note1}
\bibinfo{note}{Note that $p$ defined here differs by a factor $z_2/z_1$ with
  the definition of $p$ in the introduction.}
\end{thebibliography}

\end{document}